\documentclass[apjl]{emulateapj}
\usepackage{amssymb}
\usepackage{graphicx,graphics}
\usepackage{epsfig}
\usepackage{dcolumn}
\usepackage{bm}
\usepackage{natbib}
\usepackage{times}
\usepackage{ulem}
\def\kmm#1  {{\bf [KMM:~ #1]~}}
\def\new#1 {{\bf #1 }}
\def\cut#1 {\sout{#1} }

\newcommand{\pks}{PKS\,1830$-$21}

\newcommand{\beq}{\begin{equation}}
\newcommand{\eeq}{\end{equation}}

\newcommand{\kms}{km~s$^{-1}$}
\newcommand{\cm}{cm$^{-2}$}
\newcommand{\otwo}{O$_2$}
\newcommand{\notwo}{\ensuremath{{\rm N(O_2)}}}
\newcommand{\nhtwo}{\ensuremath{{\rm N(H_2)}}}
\newcommand{\htwo}{H$_2$}
\newcommand{\xotwo}{\ensuremath{{\rm X(O_2)}}}

\shorttitle{Constraints on the O$_2$ abundance at $z \sim 0.886$}
\shortauthors{Kanekar \& Meier}
\begin{document}
\title{A new constraint on the molecular oxygen abundance at $z \sim 0.886$}

\author{Nissim Kanekar\altaffilmark{1},
David S. Meier\altaffilmark{2}
}
\altaffiltext{1}{Swarnajayanti Fellow; National Centre for Radio Astrophysics, 
Tata Institute of Fundamental Research, Ganeshkhind, Pune - 411007, India; nkanekar@ncra.tifr.res.in}
\altaffiltext{2}{Department of Physics, New Mexico Institute of Mining and Technology, 801 Leroy Place, Socorro, NM 87801}

\begin{abstract}
We report Karl G. Jansky Very Large Array (VLA) and Atacama Large Millimeter Array (ALMA)
spectroscopy in the redshifted molecular oxygen (O$_2$) 56.265~GHz and 424.763~GHz 
transitions from the $z=0.88582$ gravitational lens towards PKS\,1830$-$21. The ALMA 
non-detection of O$_2$ 424.763~GHz absorption yields the $3\sigma$ upper limit 
$N({\rm O}_2) \leq 5.8 \times 10^{17}$~cm$^{-2}$ on the O$_2$ column density, assuming
that the O$_2$ level populations are thermalized at the gas kinetic temperature of 80~K. 
The VLA spectrum shows absorption 
by the CH$_3$CHO 56.185~GHz and 56.265~GHz lines, with the latter strongly blended with 
the O$_2$ 56.265~GHz line. Since the two CH$_3$CHO lines have the same equilibrium 
strength, we used the known CH$_3$CHO 56.185~GHz line profile to subtract out the 
CH$_3$CHO 56.265~GHz feature from the VLA spectrum, and then carried out a 
search for O$_2$ 56.265~GHz absorption in the residual spectrum. The non-detection of 
redshifted O$_2$ 56.265~GHz absorption in the CH$_3$CHO-subtracted VLA spectrum 
yields $N({\rm O}_2) \leq 2.3 \times 10^{17}$~cm$^{-2}$. Our $3\sigma$ limits 
on the O$_2$ abundance relative to H$_2$ are then $X({\rm O}_2) \leq 9.1 \times 10^{-6}$ 
(VLA) and $X({\rm O}_2) \leq 2.3 \times 10^{-5}$ (ALMA). These are $5-15$ times
lower than the best previous constraint on the O$_2$ abundance in an external galaxy. 
The low O$_2$ abundance in the $z= 0.88582$ absorber may arise due to its high neutral 
carbon abundance and the fact that its molecular clouds appear to be diffuse or 
translucent clouds with low number density and high kinetic temperature.
\end{abstract}


\maketitle
\section{Introduction} 
\label{sec:intro}

Molecular oxygen (\otwo) has long been identified as a critical species 
for the understanding of cooling and energy balance in molecular clouds, and 
of interstellar chemistry \citep[e.g.][]{goldsmith78,goldsmith11}. In standard models of 
chemistry, the \otwo\ abundance relative to that of molecular hydrogen 
${\rm H}_2$ is expected to rise to $X({\rm O}_2) \equiv 
N({\rm O}_2)/N({\rm H}_2) \sim 10^{-5}$, comparable to the carbon monoxide 
abundance, at times beyond $\sim 3 \times 10^5$~years \citep[e.g.][]{herbst73,marechal97}.
Remarkably, despite numerous searches with the {\it Submillimeter Wave Astronomy Satellite}, and the 
{\it Odin} and {\it Herschel} satellites, \otwo\ has been detected in only two 
directions in the Galaxy, towards $\rho$Oph\,A \citep{larsson07,liseau12} and Orion 
H$_2$ Peak~1 \citep{goldsmith11,chen14}, with abundances $X({\rm O}_2) \approx 5 \times 
10^{-8}$ \citep[$\rho$Oph\,A;][]{larsson07,liseau12} and $\approx 10^{-6}$ 
\citep[Orion H$_2$ Peak~1, whose relatively high abundance has been explained as arising due 
to a low-velocity C-type shock, with a modest far-ultraviolet radiation field][]
{goldsmith11,chen14,melnick15}. The majority of searches have yielded low \otwo\ 
abundances in both diffuse and dark clouds, $X({\rm O}_2) < 10^{-7}$ \citep[e.g.][]{pagani03,yildiz13}, 
two orders of magnitude lower than expected.
Although many attempts have been made to explain the paucity of \otwo\ 
\citep[e.g.][]{bergin00,charnley01,quan08,hollenbach09,whittet10}, the low \otwo\ 
abundances in molecular clouds remain a serious problem for models of chemistry. 


For cosmologically-distant galaxies, the \otwo\ lines are redshifted outside
the telluric bands and can be observed with ground-based telescopes. Unlike satellite-based 
\otwo\ {\it emission} searches, where the large telescope beam means that the derived 
$\xotwo$ is an average over multiple molecular clouds, searches for \otwo\ in {\it absorption} 
towards compact radio sources provide estimates of $\xotwo$ in individual clouds along the 
sightline. Such observations are especially interesting for high-$z$ systems as 
they allow studies of interstellar chemistry in much younger galaxies.  

The two best targets for a search for redshifted \otwo\ in absorption are 
the spiral gravitational lenses at $z \sim 0.685$ and $z \sim 0.886$ towards B0218+357 
and PKS1830$-$21, respectively, which show absorption in a variety of molecular species
\citep[e.g.][]{wiklind95,wiklind96,wiklind98,combes97,chengalur99b,kanekar03c,henkel05,muller14}. 
Molecular absorption studies of these galaxies have been used to determine 
physical conditions in the absorbing clouds \citep[e.g.][]{henkel08,menten08},
to estimate the temperature of the microwave background \citep[e.g.][]{muller13}, and even
to constrain changes in the fundamental constants of physics 
\citep[e.g.][]{kanekar11,bagdonaite13,kanekar15}. 

Searches for \otwo\ absorption have been carried out at $z = 0.685$ towards B0218+357 in the 
\otwo\,368~GHz and 424~GHz transitions \citep{combes95} and the \otwo\,56~GHz and 119~GHz transitions 
\citep{combes97b}. These yielded the upper limit ${\rm N(O_2)} < 2.9 \times 10^{18}$~\cm\ 
on the \otwo\ column density, where we have updated the results of \citet{combes97b} for an 
\otwo\ excitation temperature equal to the inferred gas kinetic temperature \citep[55~K;][]{henkel05}. 
The \htwo\ column density of the $z \sim 0.685$ absorber is $\approx 2 \times 10^{22}$~\cm\ 
\citep{gerin97,kanekar02}; this yields $X({\rm O}_2) \leq 1.5 \times 10^{-4}$, three orders 
of magnitude poorer than the limits from Galactic studies \citep[e.g.][]{pagani03}.


We have used the Karl G. Jansky Very Large Array (VLA) and the Atacama Large Millimeter Array (ALMA)
to search for redshifted \otwo\ absorption in the $z = 0.886$ spiral lens towards PKS1830$-$21. 
In this {\it Letter}, we report results from our observations, which yield stringent constraints 
on the \otwo\ abundance in this galaxy.

\begin{figure*}[t!]
\centering
\includegraphics[scale=0.4]{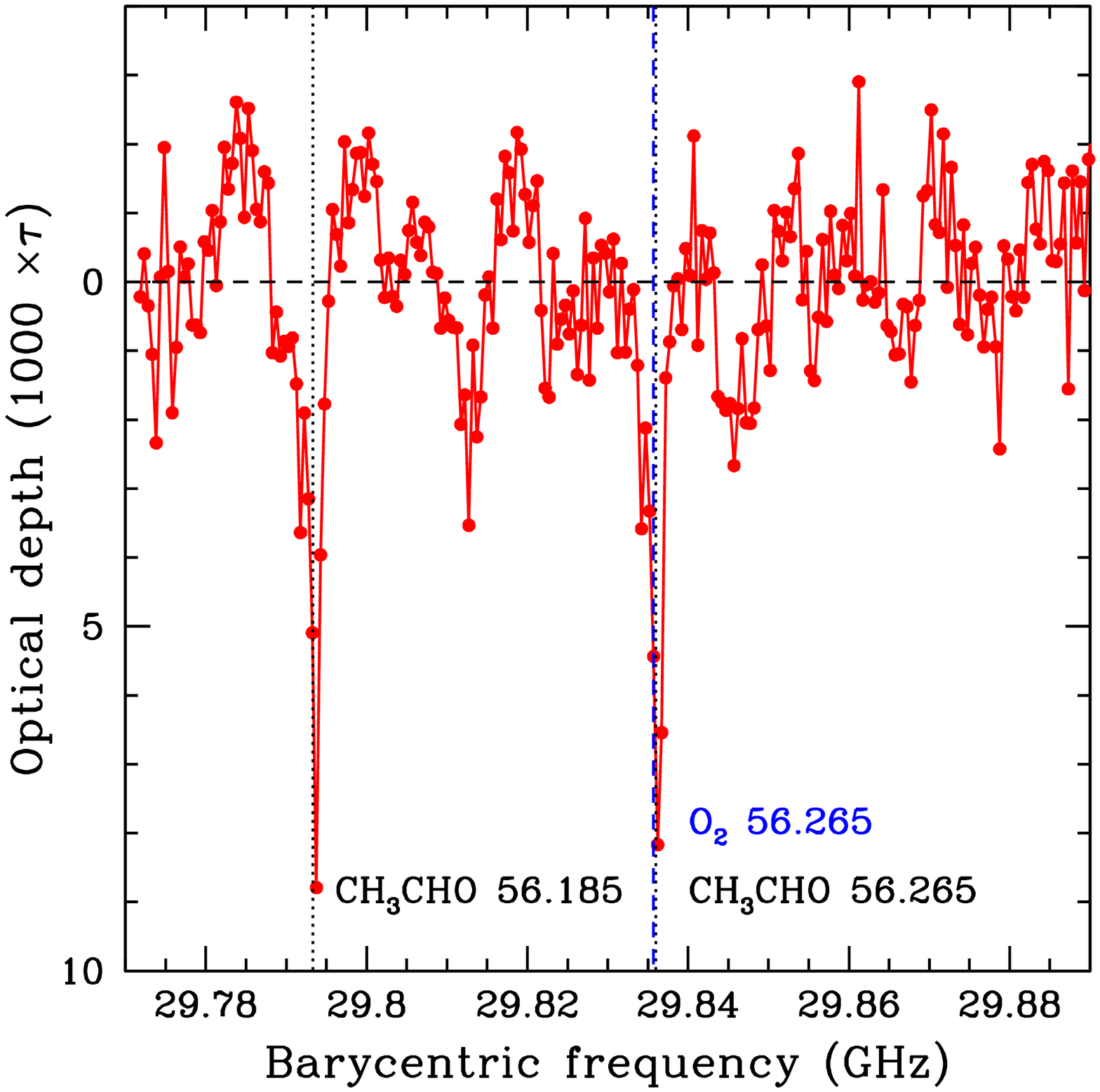}
\includegraphics[scale=0.4]{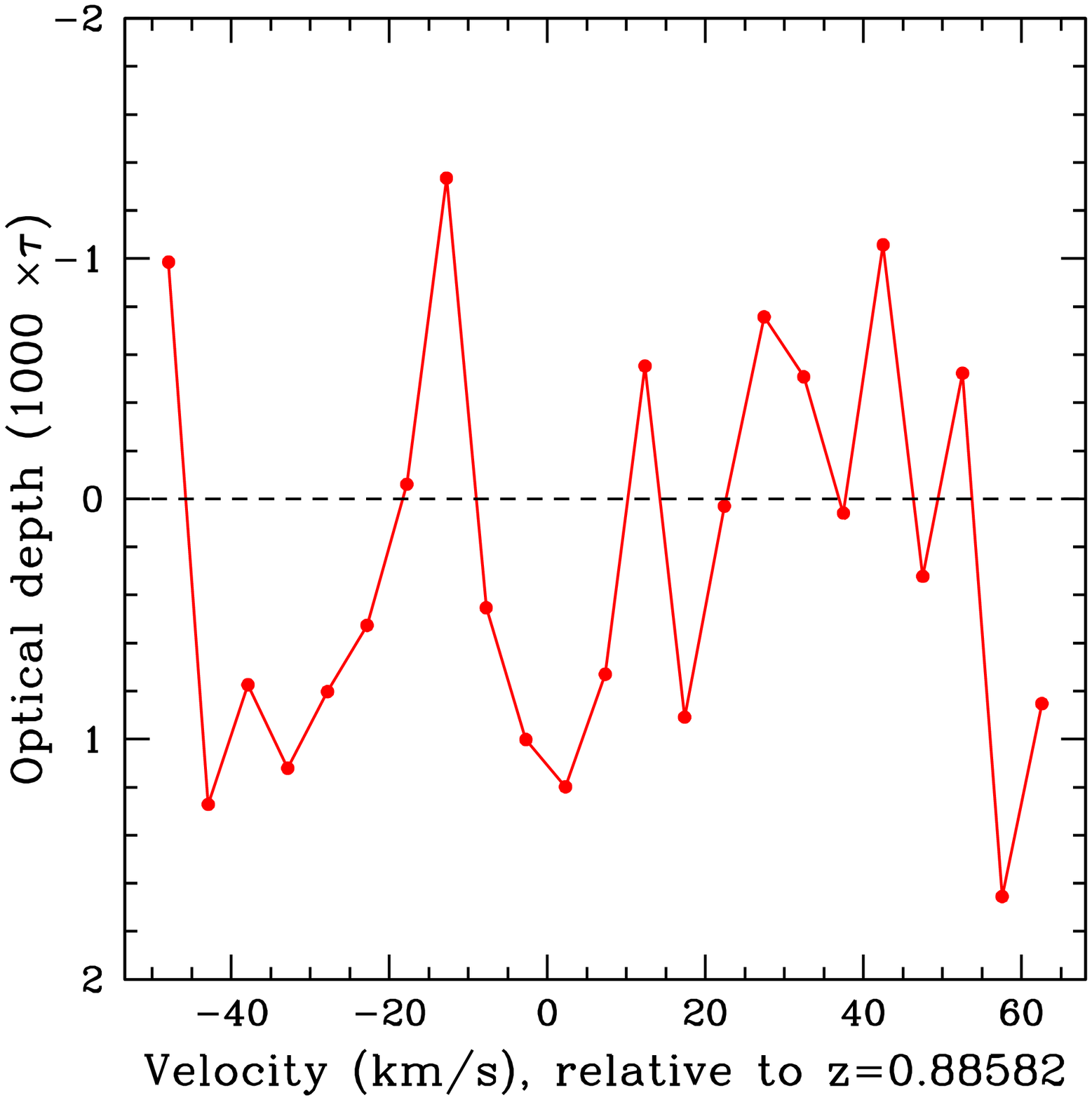}
\caption{Left panel: The final VLA spectrum towards \pks, with optical depth against the S-W 
component (in units of $1000 \times \tau$) plotted against barycentric frequency, in GHz.
The dashed vertical line indicates the redshifted \otwo~56~GHz line frequency, while the dotted
vertical lines indicate the redshifted frequencies of the two CH$_3$CHO lines (one of which is in 
excellent agreement with the \otwo\ frequency).
Right panel: The residual VLA spectrum, with optical depth plotted versus velocity (in \kms, relative to 
$z=0.88582$), covering the velocities around the redshifted \otwo~56~GHz line frequency after 
subtracting out the CH$_3$CHO~56.185~GHz line profile from the spectrum. 
\label{fig:vla}}
\end{figure*}

\section{Observations, data analysis and results}
\label{sec:data}

\begin{figure*}[t!]
\centering
\includegraphics[scale=0.4]{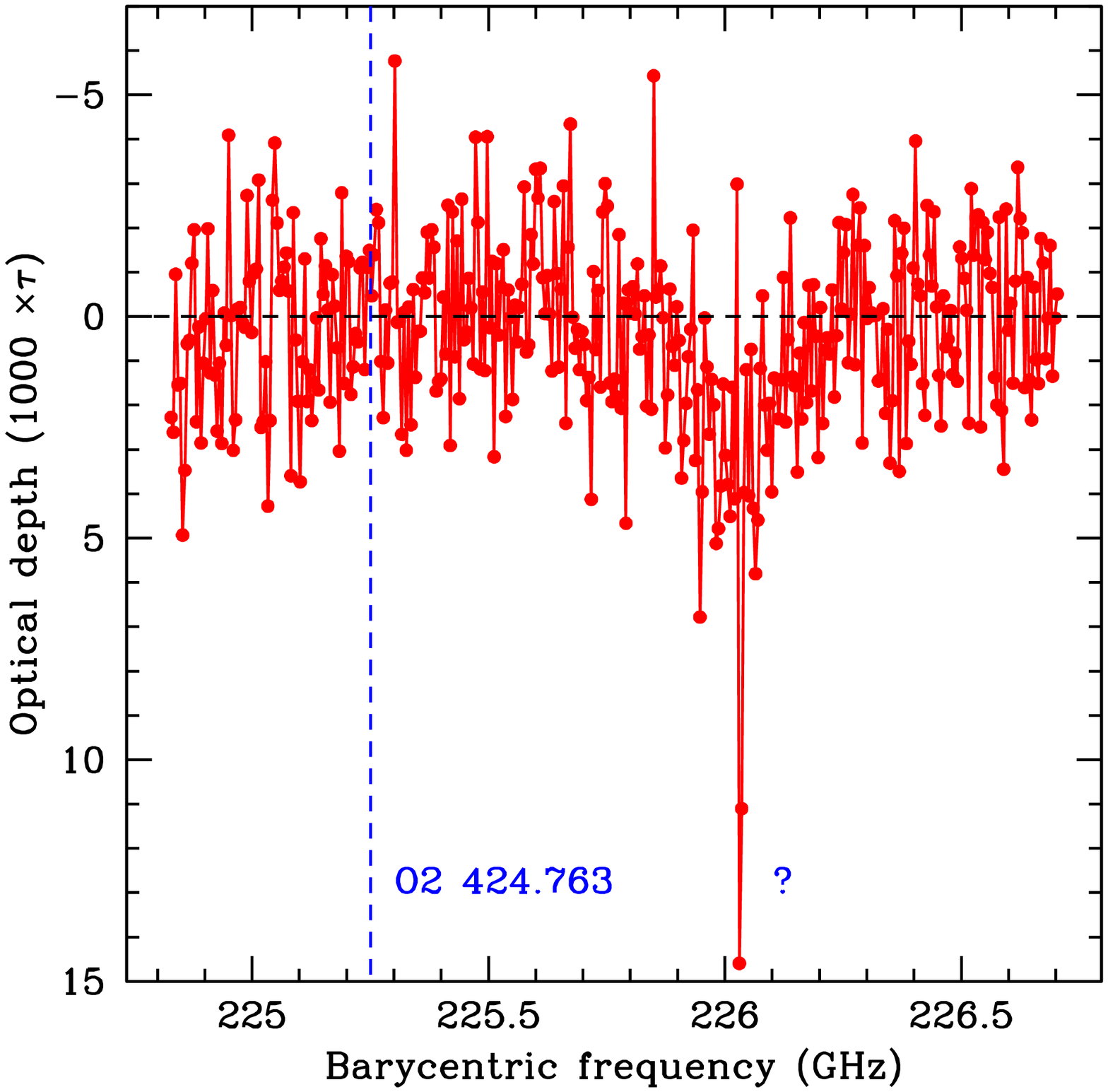} 
\includegraphics[scale=0.4]{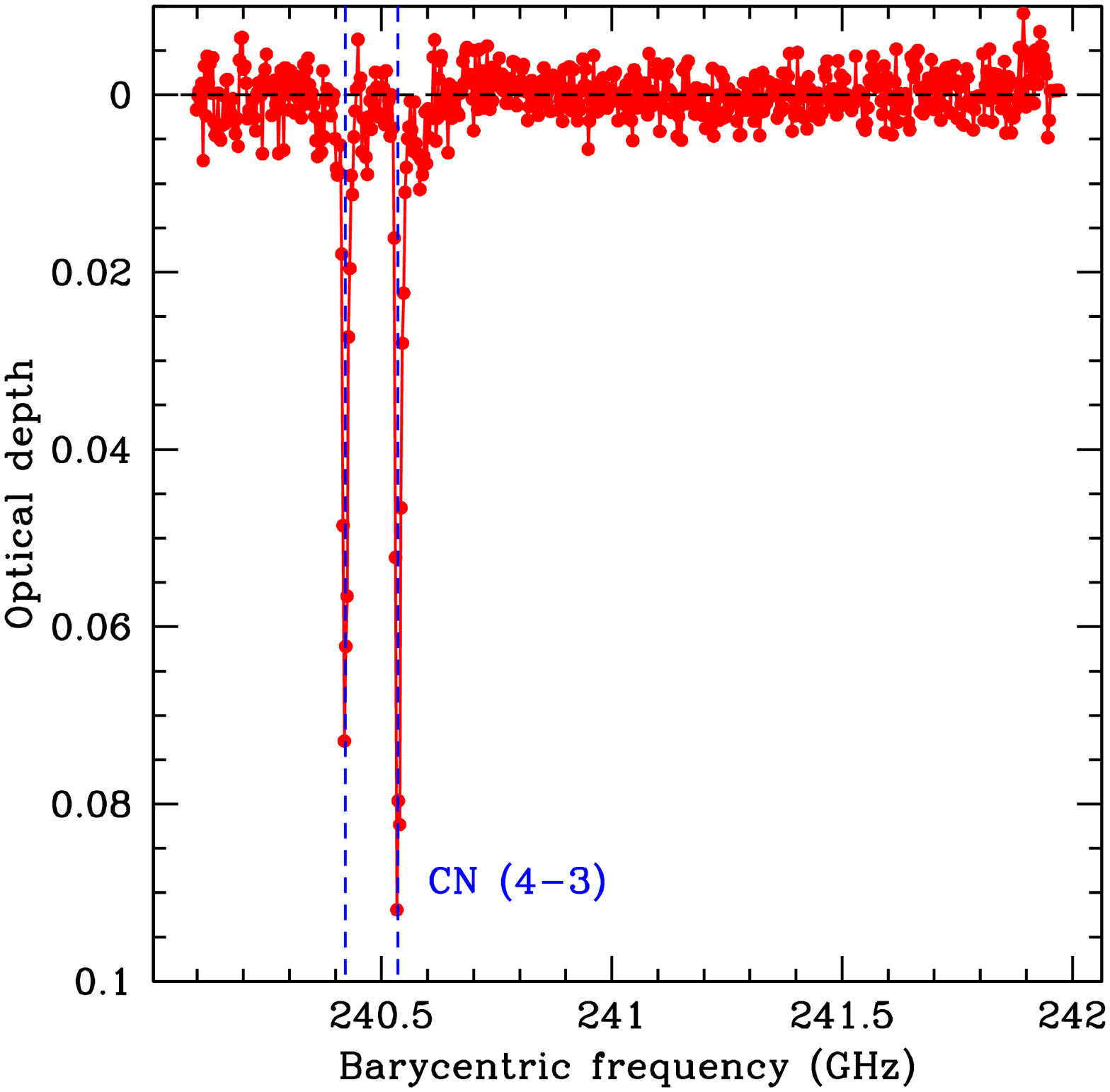} 
\includegraphics[scale=0.4]{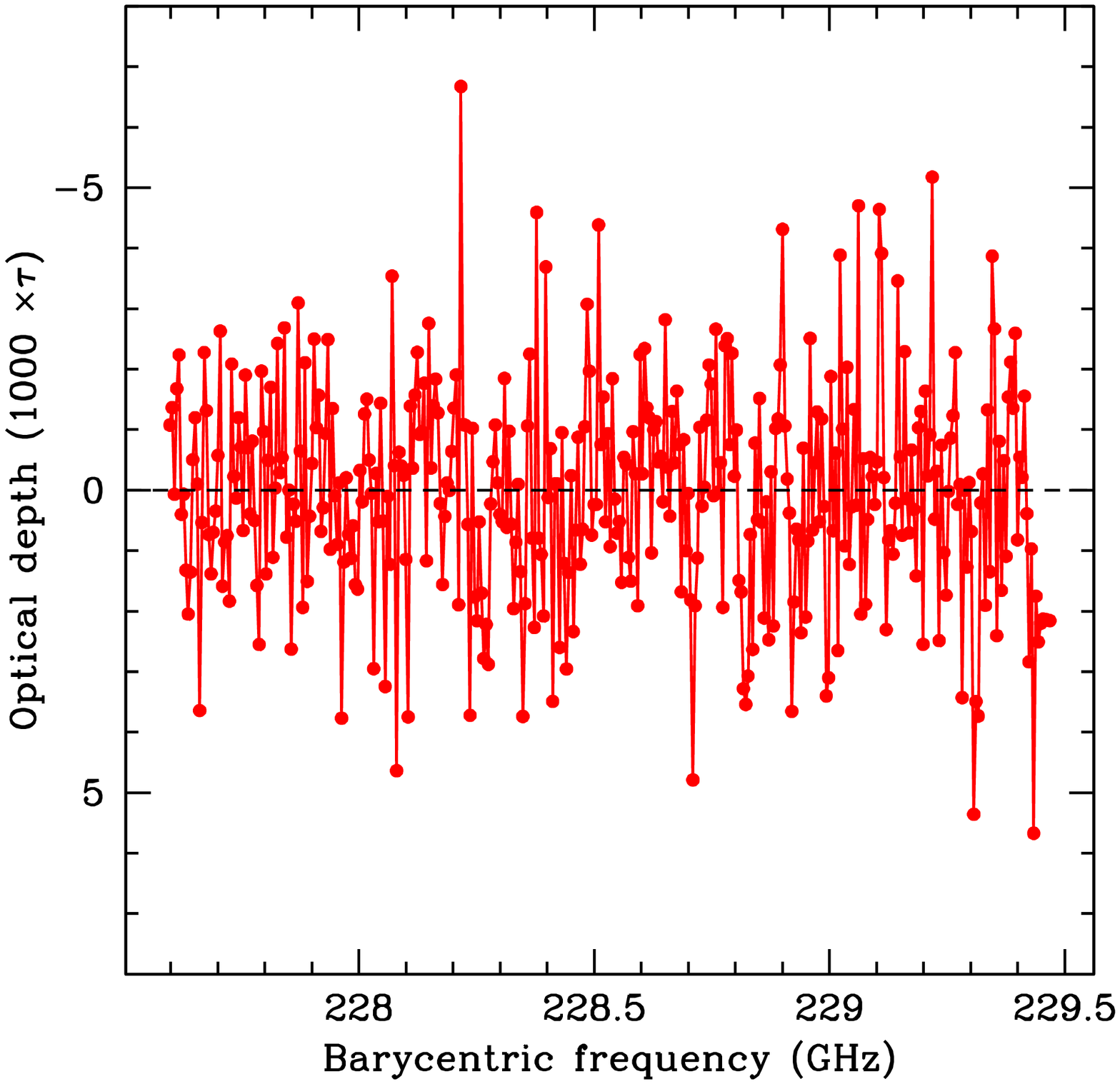} 
\includegraphics[scale=0.4]{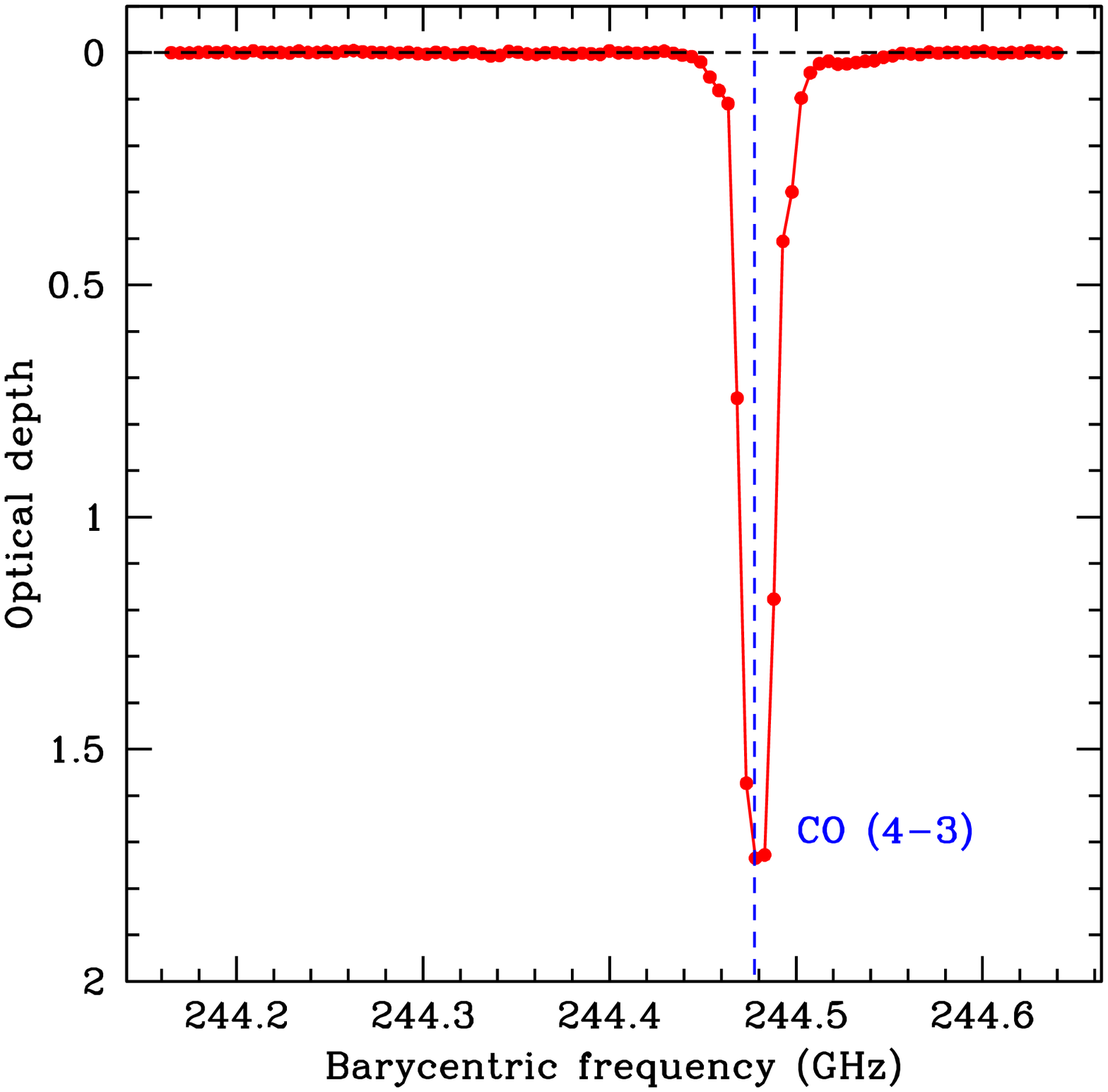}
\caption{Spectra towards \pks\ from the four ALMA IF bands, at a velocity resolution of $\approx 6.5$~\kms. 
The dashed vertical line in the top left panel indicates the redshifted \otwo~424~GHz line frequency; no 
absorption is detected here. The two absorption features in the top right panel are from the CN~(4--3)
~453.392~GHz and 453.607~GHz transitions, while the strong feature in the bottom right panel 
is the CO~(4--3) transition. The transition giving rise to the absorption feature at $\approx 226.033$~GHz 
(indicated by a question mark) in the top left panel remains unidentified.
\label{fig:alma}}
\end{figure*}

\subsection{VLA observations}

The Ka-band receivers of the VLA were used in July~2010 to carry out a search for 
\otwo\ $1_{2}\rightarrow 1_{1}$ 56.2648~GHz absorption at $z = 0.88582$ towards
\pks\ (proposal~AK725). The observations used the WIDAR correlator as the backend, with a single 
128~MHz band, sub-divided into 256 channels, centred at the redshifted \otwo\ line
frequency of 29.835~GHz, and two circular polarizations. Observations of 3C286 and the 
bright sources 3C273 and J2253+1608 were used to calibrate the flux density scale and 
the system bandpass, respectively. The total on-source time was 2~hours, with 19 working 
antennas in the VLA C-configuration

The VLA data were analysed in ``classic'' {\sc aips} using standard procedures. Note that \pks\ 
is unresolved by our 19-antenna VLA C-array at Ka-band. After initial calibration, the tasks {\sc uvsub}
and {\sc uvlin} were used to subtract the image of \pks\ from the calibrated visibilities, 
and then to subtract out any residual continuum by fitting a linear baseline to line-free
channels. The residual visibilities were then imaged and the final spectrum covering the 
redshifted \otwo~56.265~GHz transition obtained by taking a cut through the spectral cube at 
the location of \pks. 

The final VLA spectrum is shown in the left panel of Fig.~\ref{fig:vla}, with optical depth against the 
S-W image component of \pks\ plotted versus heliocentric frequency, in GHz. The root-mean-square 
(RMS) noise on the spectrum is $\approx 1.1 \times 10^{-3}$ per 5~\kms\ channel, in optical 
depth units \citep[assuming that the S-W component contains $\approx 38$\% of the total flux density 
of \pks\ at these frequencies; e.g.][]{muller11}. A strong absorption feature, with 
an integrated optical depth of $\approx (0.146 \pm 0.014)$~\kms, is clearly 
visible at the expected frequency of the redshifted \otwo~56.265~GHz line (indicated by the dashed 
vertical line). However, it was realized that there is a CH$_3$CHO transition ($3_{-1,3} 
\rightarrow 2_{-1,2}$~E) at a rest frequency of 56.2652~GHz that would be strongly 
blended with the \otwo~56.265~GHz line, and that might cause the observed absorption. 
Further, a second absorption feature is visible at $\approx 29.793$~GHz, which could 
be redshifted CH$_3$CHO absorption, in the $3_{1,3} \rightarrow 
2_{1,2}$~A++ transition. If the two features indeed arise from CH$_3$CHO, it would be difficult 
to draw conclusions about the \otwo\ abundance (although see below). We hence carried out an 
ALMA search for redshifted \otwo~424~GHz absorption, to test whether the VLA absorption feature 
indeed arises from the \otwo~56.265~GHz line.

\subsection{ALMA observations}

The Band-6 receivers of ALMA were used in March~2014 to search for redshifted 
\otwo\ $1_{2}\rightarrow 3_{2}$ 424.7631~GHz absorption at $z=0.88582$ towards \pks. 
The observations used four 1.875~GHz intermediate frequency (IF) bands, each sub-divided 
into 3840 channels, and with 2 polarizations. The four IF bands were centred at 225.780~GHz 
(covering the redshifted \otwo\ 424~GHz line frequency), 228.530~GHz, 241.033~GHz and 243.733~GHz.
Observations of Titan, J1733$-$1304, J1923$-$2104, and a few calibrators were used to 
calibrate the flux density scale and the system bandpass and gain. The total on-source time 
was $\approx 2$~hours, with 25 ALMA antennas.

The ALMA data were analysed in two stages, first using the {\sc casa} pipeline to carry 
out the initial calibration procedure, and then self-calibrating the data of \pks\ in 
{\sc aips}. The flux density scale was calibrated using the short-baseline data 
on Titan, and this was then extended to longer baselines by bootstrapping the data of 
J1923$-$2104. The data of J1733$-$1304 and J1923$-$2104 were, respectively, used to 
calibrate the system bandpass and initial gain. After applying the initial calibration in
{\sc casa}, a standard self-calibration procedure was used in {\sc aips}, with a few rounds 
of phase-only self-calibration followed by a single round of amplitude-and-phase
self-calibration. The final image has a synthesized beam of $\approx 1.0'' \times 0.8''$
(with the two strong image components of \pks\ marginally resolved), and an RMS noise 
of $\approx 0.14$~mJy/Beam. The task {\sc jmfit} was used to measure the flux densities of 
the N-E and S-W image components, via a 2-Gaussian fit to the final image; this yielded flux 
densities of $549.98 \pm 0.43$~mJy (N-E) and $342.63 \pm 0.43$~mJy (S-W). The continuum 
image of \pks\ was then subtracted from the calibrated visibilities of each IF band using 
the task {\sc uvsub}, and the residual visibilities of each band were then imaged to produce 
a spectral cube, after shifting the data to the heliocentric frame. The spectrum for each IF band was 
then produced via a cut through the cube at the location of the S-W image component. The final spectra 
have an RMS noise of $\approx 1.0-1.3$~mJy at the re-sampled velocity resolution of $\approx 1.3$~km/s. 

The final Hanning-smoothed and re-sampled spectra from the four ALMA IF bands (after 
subtracting a second-order baseline) against the S-W component are shown in the four panels of 
Fig.~\ref{fig:alma}, with optical depth plotted versus heliocentric frequency, in GHz. All 
spectra are shown after smoothing to, and re-sampling at, a velocity resolution of 
$\approx 6.5$~\kms\, the resolution at which the search for redshifted \otwo\ 424~GHz absorption 
was carried out. No evidence for \otwo\ 424~GHz absorption can be discerned in the spectrum in the 
top left panel of Fig.~\ref{fig:alma}. The final RMS noise on the spectrum is $\approx 
1.8 \times 10^{-3}$ per 6.5~\kms\ channel, in optical depth units.

In passing, we note that four absorption features were clearly detected in the ALMA spectra;
three of these correspond to the CO~(4--3) and two CN~(4--3) transitions (see Fig.~\ref{fig:alma}).
However, we have been unable to identify the fourth transition, at $\approx 226.033$~GHz, 
i.e. at rest-frame frequencies of 426.257~GHz (at $z=0.88582$, the absorber being studied here), 
792.698~GHz \citep[at $z=2.507$, the redshift of \pks;][]{lidman99} or 269.567~GHz 
\citep[at $z=0.1926$, the redshift of another known absorber towards \pks;][]{lovell96}. The 
line width is $\approx 5$~\kms, similar to that of other high-frequency transitions from the 
$z = 0.88582$ absorber. It appears that this is not a known low-energy transition of a species 
expected to be abundant in the ISM.

\section{Discussion}
\label{sec:alpha}

\begin{figure}[t!]
\centering
\includegraphics[scale=0.4]{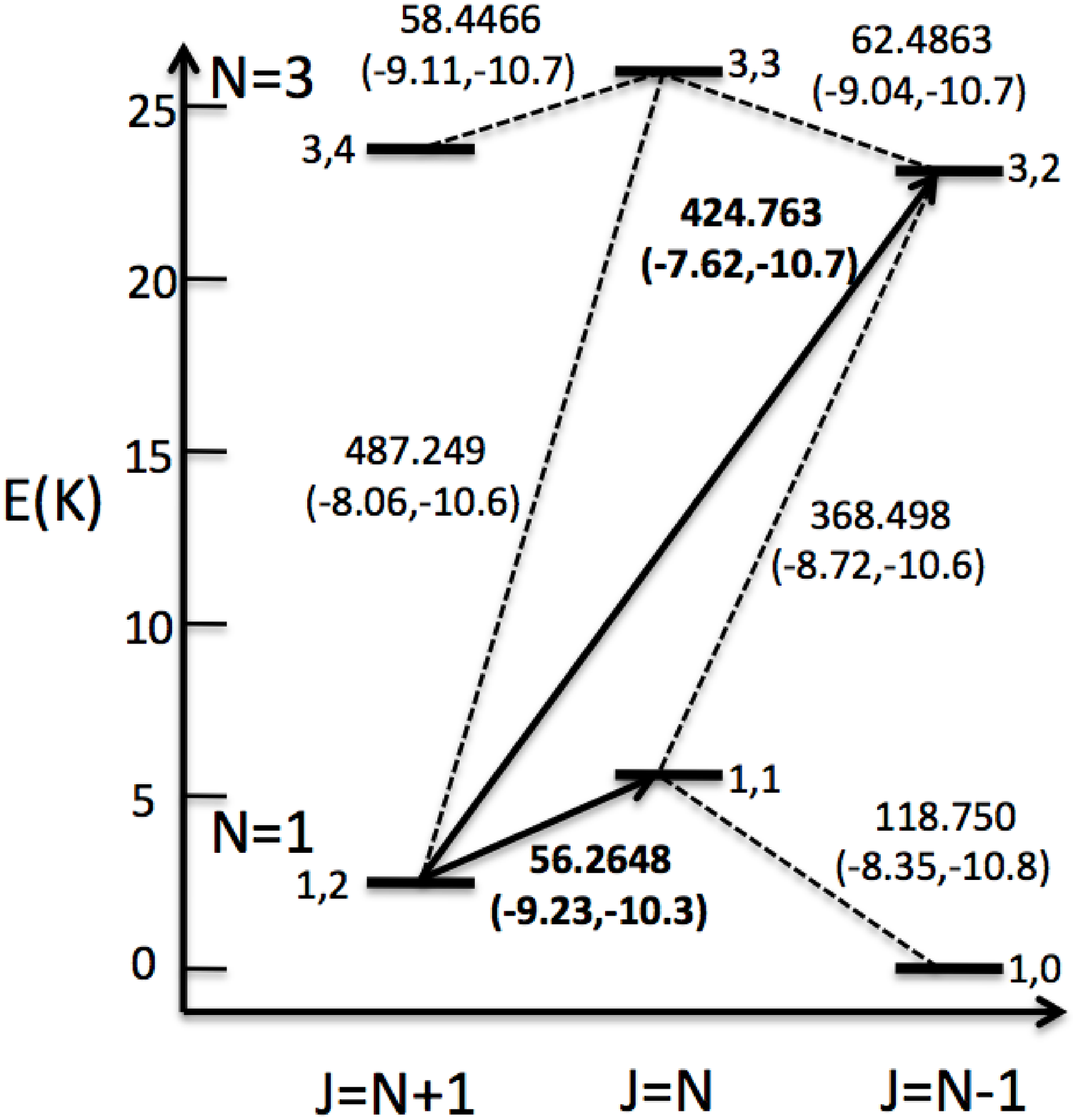}
\caption{Low energy level diagram for \otwo, adapted from \citet{marechal97},
with levels labeled by their (N,J) quantum numbers. The observed transitions 
and transitions of comparable energy are displayed with solid and dashed arrows, 
respectively. Transition are labeled by their rest frequency \citep[in GHz, 
from Splatalogue; e.g.][]{pickett98,remijan07,drouin10} and radiative/collisional rate coefficients
[as (${\rm log}[A_{ul}],{\rm log}[C_{ul}]$), with values for $C_{ul}$ from the RADEX database; \citealp{schoier05,lique10}].
\label{fig:otwo}}
\end{figure}

The first question is whether the absorption feature seen at $\approx 29.836$~GHz in the 
VLA spectrum of \pks\ arises from \otwo\ or from CH$_3$CHO (or, indeed, some other transition).
The lower energy level of the \otwo\ 56~GHz and \otwo\ 424~GHz transitions is the 
same (the $1_{2}$ state; see Fig.~\ref{fig:otwo}), permitting a direct comparison between the expected optical depths 
in the two lines. Of course, the ratio of the line strengths depends on the respective 
excitation temperatures. In the case of the $z = 0.88582$ absorber, the number density $n_{\rm H_2}$
and kinetic temperature $T_k$ of the molecular gas have been estimated to be 
$n_{\rm H_2} \sim 1700-2600$~cm$^{-3}$ and $T_k \approx 80$~K \citep{henkel08,henkel09}. 
For number densities $\gtrsim10^3$~cm$^{-3}$ and $T_k \geq 30$~K, the \otwo\ line populations 
are expected to be thermalized \citep[e.g.][]{goldsmith00}, i.e. $T_x \approx T_k$. For 
$T_x \approx 80$~K, the 424~GHz line is expected to be slightly stronger than the 
56~GHz line, $\tau_{424} \approx 1.1 \times \tau_{56}$. Our ALMA $3\sigma$ limit on 
the integrated \otwo\ 424~GHz optical depth is $\approx 0.037$~\kms, a factor of 5 lower than 
the integrated optical depth ($\approx 0.146 \pm 0.014$~\kms) of the 29.836~GHz absorption feature 
in the VLA spectrum. We can thus conclusively rule out the possibility that the VLA absorption 
feature arises from the \otwo\ 56~GHz transition. The feature is most likely to arise from 
the CH$_3$CHO $3_{-1,3} \rightarrow 2_{-1,2}$\,E transition.

The ALMA upper limit to the \otwo\ 424~GHz optical depth can be used to place a limit on 
the total \otwo\ column density. For $T_x = 80$~K, this gives $\notwo \leq 5.8 
\times 10^{17}$~\cm, at $3\sigma$ significance. 

Interestingly, the two CH$_3$CHO transitions ($3_{-1,3} \rightarrow 2_{-1,2}$~E and 
$3_{1,3} \rightarrow 2_{1,2}$~A$++$) seen in the VLA spectrum at, respectively, 29.836~GHz 
and 29.793~GHz, have the same line strengths. One can hence subtract one from the other 
to search for any additional absorption arising from the \otwo~56~GHz line. This was done 
by using two-point interpolation to resample the CH$_3$CHO $3_{1,3} \rightarrow 2_{1,2}$~A$++$
line profile at the measured velocities of the CH$_3$CHO $3_{-1,3} \rightarrow 2_{-1,2}$~E 
line, and then subtracting out the resampled line profile from the latter spectrum. This procedure
is unlikely to yield any systematic effects, as both CH$_3$CHO line profiles are well-sampled, with 
at least 3 independent spectral points detected at $> 5\sigma$ significance. No absorption is 
detected in the residual VLA spectrum, yielding an integrated \otwo~56~GHz optical depth of 
$< 0.0131$~\kms, again at $3\sigma$ significance, against the S-W 
component of \pks. Again using $T_x = 80$~K, this yields $\notwo \leq 2.3 \times 10^{17}$~\cm, 
a factor of $\approx 2.5$ more stringent than the ALMA upper limit.

The \htwo\ column density of the $z \sim 0.886$ lens has been estimated 
to be $\approx 2.5 \times 10^{22}$~\cm\ \citep{gerin97,wiklind98}. These are broadly 
consistent with estimates of the total hydrogen column density towards both 
lensed images from Chandra 
and ROSAT X-ray spectroscopy, N(H)~$= (1.8-3.5) \times 10^{22}$~\cm\ \citep{mathur97,dai06}.
Note that, while \citet{muller08} argue that the \htwo\ column density may be an 
order of magnitude larger than the above values to account for the detection of species 
such as HC$^{17}$O$^+$ and HC$^{15}$N in absorption, such a high value of $\nhtwo$ appears 
to be ruled out by the X-ray data. Using a value of $\nhtwo = 2.5 \times 10^{22}$~\cm\ 
yields \otwo\ abundances of $X({\rm O}_2) \leq 9.1 \times 10^{-6}$ and 
$X({\rm O}_2) \leq 2.3 \times 10^{-5}$ from the VLA (CH$_3$CHO-subtracted) \otwo~56~GHz 
and the ALMA \otwo~424~GHz non-detections, respectively. Of course, if the higher 
\htwo\ column density estimate of \citet{muller11} is correct, then our constraints 
on the \otwo\ abundance would be more stringent by an order of magnitude, i.e. 
$X({\rm O}_2) \leq 9.1 \times 10^{-7}$.

Prior to this work, the strongest constraint on the \otwo\ abundance outside the Milky 
Way was $X({\rm O}_2) \leq 1.5 \times 10^{-4}$ in the $z = 0.685$ absorber towards B0218+357
\citep[see Section~1;][]{combes97b}. Our VLA upper limit on the \otwo\ abundance in the $z = 0.886$ absorber 
towards \pks\ is a factor of $\approx 15$ lower than this, and comparable to the measured
\otwo\ abundance towards the Orion H$_2$ Peak~1 \citep{goldsmith11}. However, our limit is nearly 
two orders of magnitude weaker than the constraints on, or measurements of, \otwo\ abundances 
in the Milky Way \citep[e.g.][]{pagani03,larsson07,liseau12}. Unfortunately, the high gas 
kinetic temperatures in the two gravitational lenses, $\approx 55$~K in the $z= 0.685$ 
absorber and $\approx 80$~K in the $z= 0.886$ absorber, imply that it will not be easy to 
improve upon our present constraint and achieve an \otwo\ abundance sensitivity comparable to 
those in the Milky Way.

While our \otwo\ abundance constraints for the $z=0.88582$ absorber are less stringent than 
those in the Galaxy, these are by far the most sensitive constraints in an external galaxy. 
Further, the Galactic estimates stem from emission studies with differing angular resolution in 
the \otwo\ and CO lines. The derived abundances are hence an average over multiple molecular 
clouds with different excitation conditions; this can imply large uncertainties in $X({\rm O}_2)$,
of upto two orders of magnitude \citep[e.g.][]{liseau10}. The resolution of the present 
interferometric absorption study is determined by the size of the background radio continuum at 
the observing frequency. For \pks, the emission from the S-W image at high frequencies 
($14.5-43$~GHz) arises in a compact source of size~$< 0.5$~mas \citep{jin03,sato13}, i.e. 
transverse size $< 4$~pc at $z = 0.88582$. The \otwo\ abundance estimates are hence likely 
to be reliable here, as both the \otwo\ and the \htwo\ column densities are inferred 
from absorption studies probing the same pencil beam towards the S-W image. 

Finally, it is clear that we rule out \otwo\ abundances of $\approx 10^{-5}$ at 
$3\sigma$ significance in the $z = 0.886$ lens towards \pks. As noted earlier, the 
\otwo\ abundance is expected to reach about this level, comparable to the CO abundance, 
in standard models of molecular chemistry within $\approx 3 \times 10^5$~years 
\citep[e.g.][]{herbst73,marechal97}. The low \otwo\ abundance thus appears a conundrum, 
even for the $z = 0.886$ absorber. A possible explanation lies in the high derived 
abundance of neutral carbon in this system by \citet{bottinelli09}, who obtain 
N(C)/N(\htwo)~$\approx 10^{-4}$, somewhat larger than the CO 
abundance. In typical molecular 
clouds, \otwo\ is destroyed by the reactions ${\rm C} + {\rm O}_2 \rightarrow {\rm CO} + {\rm O}$, 
${\rm C}^+ + {\rm O}_2 \rightarrow {\rm CO} + {\rm O}^+$, and 
${\rm C}^+ + {\rm O}_2 \rightarrow {\rm CO}^+ + {\rm O}$. The high carbon abundance 
in the $z = 0.886$ absorber is thus unfavourable for the survival of \otwo, and can account
for its low abundance. \citet{bottinelli09} also note that the high carbon abundance 
relative to CO suggests that the absorbing gas arises in translucent clouds, or 
clouds in an early phase of the transition from diffuse to dense gas, with low 
densities and mild-UV fields. This is consistent with the high gas kinetic temperature,
($\approx 80$~K), and relatively low densities ($\approx 1700 - 2600$~cm$^{-3}$) obtained
by \citet{henkel09}, implying that the absorber at $z = 0.88582$ does not arise 
in a classical dark cloud.

In summary, we have used the VLA and ALMA to obtain tight constraints on the \otwo\ 
abundance (relative to \htwo), $\xotwo \leq 9.1 \times 10^{-6}$, in the $z= 0.88582$ 
spiral gravitational lens towards \pks. This is a factor of $\gtrsim 15$ more stringent than the 
best previous constraint on the \otwo\ abundance in an external galaxy. We argue that 
the low \otwo\ abundance in the $z \sim 0.886$ lens may arise due to its high neutral 
carbon abundance (resulting in the efficient destruction of \otwo), and the fact 
that the absorbing clouds are probably not dark clouds, but instead diffuse or 
translucent clouds, with relatively low number density and high gas kinetic temperature.

%

\acknowledgments

This paper makes use of the following ALMA data: ADS/JAO.ALMA\#2012.1.00581.S. ALMA is a 
partnership of ESO (representing its member states), NSF (USA) and NINS (Japan), together 
with NRC (Canada), NSC and ASIAA (Taiwan), and KASI (Republic of Korea), in cooperation 
with the Republic of Chile. The Joint ALMA Observatory is operated by ESO, AUI/NRAO and 
NAOJ. This paper also makes use of VLA data (proposal 10A-110). The National Radio 
Astronomy Observatory is operated by Associated Universities, Inc, under cooperative 
agreement with the NSF. NK acknowledges support from the Department of Science and Technology 
via a Swarnajayanti Fellowship. DSM acknowledges partial support by the National Science Foundation
through grant AST-1009620.

\bibliographystyle{apj}

\end{document}